\documentclass{emulateapj}

\def\eqq#1{Equation~(\ref{#1})}
\newcommand\etal{{\it et al.\/}}
\newcommand\eg{{\it e.g.\/}}
\newcommand\ie{{\it i.e.\/}}
\newcommand{\bfx}{\mbox{\bf x}}
\newcommand{\bfe}{\mbox{\bf e}}
\newcommand{\bfd}{\mbox{$\bf d$}}
\newcommand{\hbfd}{\mbox{$\bf \hat d$}}

\begin{document}

\journalinfo{CVS \$Revision: 2.2 $ $ \$Date: 2003/09/11 13:30:28 $ $}
\submitted{accepted to ApJ}

\title{Dark Energy Constraints from Weak Lensing Cross-Correlation
Cosmography} 

\author{G. Bernstein \& B. Jain}
\affil{Dept. of Physics and Astronomy, University of Pennsylvania,
Philadelphia, PA 19104}
\email{garyb, bjain@physics.upenn.edu}

\begin{abstract}
We present a method to implement the idea of \citet{JT} to constrain
cosmological parameters with weak gravitational lensing. 
Photometric redshift information on foreground galaxies is used to
produce templates of the mass structure at foreground slices $z_\ell$,
and the predicted distortion field is cross-correlated with the
measured shapes of sources at redshift $z_s$.  The variation of the
cross-correlation with $z_s$ depends purely on ratios of angular
diameter distances.  We propose a formalism for such an analysis that
makes use of all foreground-background redshift pairs, and derive the
Fisher uncertainties on the dark energy parameters that would result
from such a survey.  Surveys from the proposed SNAP satellite or
the LSST observatory could constrain the dark energy equation of state to
$\sigma_{w_0}\approx 0.01 f_{\rm sky}^{-1/2}$ and 
$\sigma_{w_a}\approx 0.035 f_{\rm sky}^{-1/2}$ after application of a
practical prior on $\Omega_m$.  Advantages of this method over
power-spectrum measurements are that it is unaffected by residual PSF
distortions, is not limited by sample-variance, and can use non-linear
mass structures to constrain cosmology.  The signal is, however, very
small, amounting to a change of a few parts in $10^3$ of the lensing
distortion.  In order to realize the full sensitivity to cosmological
parameters, the calibration of lensing distortion must be independent
of redshift to comparable levels, and photometric redshifts must be
similarly free of bias.  
Both of these tasks require substantial advance over the present state
of the art, but we discuss how such
accurate calibrations might be achieved using internal consistency tests.
Elimination of redshift bias would
require spectroscopic redshifts of $\sim 10^4-10^5$ high redshift
galaxies---fewer for lensing surveys less ambitious than SNAP or LSST. 

\end{abstract}


\section{Introduction}
Weak gravitational lensing is already one of the more accurate
constraints on cosmological parameters: several groups have measured
the power spectrum of the shear induced on background galaxies by
foreground mass fluctuations, leading to $\approx10\%$ constraints on
the matter power spectrum normalization $\sigma_8$
\citep{Bacon,Brown,Hamana02,Hoek02,Mike03,RRG02,vW02}.
The weak lensing method has potential for much more precise constraints on
$\sigma_8$ and other parameter combinations, once larger sky areas are
surveyed.  The addition of redshift information on the source galaxies
allows measurement of the mass power spectrum as a function of
redshift, which is expected to greatly increase the accuracy of weak
lensing cosmological constraints
\citep{Hu99,Hu02a,Hu02b,Huterer02,Kev02,Heavens03,Refregier03,Knox03}
and permit reconstruction of the three-dimensional mass distribution
\citep{Taylor,HK02}. 
Weak lensing surveys currently underway such as the CFHT
Legacy Survey\footnote{\tt www.cfht.hawaii.edu/Science/CFHLS/} (CFHLS) and the
Deep Lens Survey\footnote{\tt dls.bell-labs.com/} will gather 
photometric redshift information to facilitate this ``tomographic'' analysis.

The successful constraint of the mass power spectrum and cosmological
parameters at $\approx1\%$ levels will require reduction of several
sources of systematic error.  On the theoretical side, there are not
yet predictions of the mass power spectrum $P(k,z)$ that are accurate
to the percent level in the non-linear regime (see \citet{LJ03} for
further discussion).  A concerted
application of $N$-body computing would likely yield the dark-matter
spectrum to desired accuracy, though on very small scales the
contribution of baryons to the power spectrum must be included, and
will be difficult to calculate.  The inaccuracies of power-spectrum
estimation may be bypassed by using only large-scale information,
where linear or perturbative calculations suffice, but this means
discarding most of the lensing information, which lies at non-linear scales.

On the measurement side, currently published power-spectrum
measurements all show contamination by systematic errors at the
$\approx10\%$ level or higher.  This systematic power is likely
residual from the process of correcting galaxy shapes for point-spread
function (PSF) effects.  New methodologies have been introduced
\citep{BJ02, Ref, K00} which should greatly reduce the
systematic contamination, but this remains to be demonstrated.  There
are also subtle difficulties in calibrating the weak lensing signal
to 1\% accuracy \citep{BJ02, Hirata}.

\citet{JT} introduce a new method for analysing
weak-lensing data with depth information
which promises to largely bypass these
systematic difficulties.  The basic concept is to use the survey's
photometric redshift data to create a map of the foreground galaxies,
from which an estimated map of the foreground mass can be made.  This
foreground mass slice induces shear on all the galaxies in the
background.  The measurement consists of tracking the amplitude of the
induced shear as a function of the background redshift.  Once a
foreground shear ``template'' at $z_l$ is created, the dependence of shear on
the source redshift $z_s$ is given solely by geometric factors.  By
taking ratios of induced shears at different $z_s$, any sensitivity to
errors in the foreground mass template is cancelled out, leaving
us with a purely geometric observable.  One is not attempting to
discern the line-of-sight structure, so this is not tomography.  We
could call this ``cross-correlation cosmography'' since the measured
quantities are metric elements of the homogeneous cosmology.

The idea of
using lensing effects on sources at differing redshift to constrain
cosmological parameters has been examined before \citep{LP98, GFM, GKS,
S02}, with Jain \& Taylor offering a methodology for massive datasets
that is insensitive to the nature of cluster or halo profiles and
examines a time-dependent dark energy component.

The practical advantages of cross-correlation cosmography include:
\begin{itemize}
\item No theoretical estimate of the mass power spectrum is required.
  The mass fluctuations are estimated directly from the foreground
  galaxy distribution, and any inaccuracies (\eg\ bias) are cancelled
  out in the analysis.  The observables are hence calculable to
  arbitrary accuracy.
\item With no need for a modelled power spectrum, we can use shear
  information on all scales sampled by the data.  The lensing
  distortion variance $\langle d^2 \rangle$ is $\sim(5\%)^2$,
  which is at least two orders of magnitude larger than the variance
  due solely to linear-regime fluctuations.  Hence the signal-to-noise
  ratio ($S/N$) of the weak lensing data will be much higher.
\item Because the background shear is being cross-correlated with a
  template, the shear signal enters only linearly into the
  statistics.  Systematic power (\eg\ PSF residuals) average to zero
  since they will not correlate with the foreground shear template.
  This makes the PSF-correction task immensely easier than for
  power-spectrum tomography, in which systematic power adds to the
  shear power.  For cross-correlation cosmography, systematic power may
  increase uncertainties but does not bias the results.  In the
language of experimenters, we have changed from a total-power
measurement to a phase-sensitive method.
\item Uncertainties in calibration of the shear signal also cancel if
  they are independent of $z_s$, since we will be interested only in
  ratios of shear at different $z_s$.  We shall see below, however,
that the method is highly sensitive to {\em differential} calibration
errors. 
\end{itemize}
Taking the ratio of shear signals at different $z_s$ provides a pure
geometric measurement, but taking this ratio does decrease the
dependence of the signal on cosmological parameters.  Large numbers of
galaxies must be surveyed in order to reduce random errors to make up
for the smaller signals, but surveys of up to $10^9$ galaxies are
currently being planned.

In this paper we investigate the potential of cross-correlation
cosmography by proposing an analysis methodolgy that appears close to
optimal, and calculating the expected cosmological-parameter
uncertainties with the Fisher-matrix method.  We then apply this
analysis to some planned weak lensing surveys.  Finally we return to
the issue of systematic errors, exploring the accuracies that will be
required for photometric redshift estimates and shear calibration.

\section{Cross-Correlation Formalism}
\subsection{The Observable Quantity}
The underlying assumption of the method is that the observed ellipticity
of background galaxy $k\in\{1,2,\ldots,N\}$ is
\begin{equation}
\label{bfdk}
\bfe^{\rm obs}_k =
\sum_\ell \bfd_\ell(\bfx_k) g_{\ell k} + \bfd_{k,\rm sys} + \bfe_k,
\end{equation}
The quantities of interest will be the geometric factors
\begin{equation}
\label{gls}
g_{\ell s} \equiv { r(\chi_s - \chi_\ell) \over r(\chi_s) },
\end{equation}
where $\chi_\ell$ and $\chi_s$ are the comoving distances to a lens
and source planes, and $r(\chi)$ is comoving angular diameter
distance.  For a flat Universe, $r(\chi)=\chi$.  Each of the
$\bfd_\ell$ is the distortion field\footnote{
The distortion denoted as {\boldmath
$\delta$} in \citet{BJ02} is labelled \bfd\ here
in order to avoid confusion with the Kronecker delta.}
imparted on a source plane at
$\chi=\infty$ by the mass in redshift shell $\ell$.  It can be
expressed as an integral over the mass distribution in the redshift
shell by well-known formulae, \eg\ \citet{BS01}.  Note that the
distortion $\bfd$ is twice the shear \mbox{\boldmath $\gamma$} in the
weak-lensing limit. 
Its variance $\langle
d^2_\ell \rangle$ can be calculated as an integral of the mass
power spectrum  $P(k,z_\ell)$, with the formula in
Appendix~\ref{pspec}.  Our method, however, will be to make 
as little use as possible of the power spectrum.

\eqq{bfdk} is correct only in the limit that the induced distortion
$\bfd_\ell$, the instrumental systematic distortion $\bfd_{k,\rm
  sys}$, and the intrinsic galaxy shape $\bfe_k$ are all $\ll 1$.  The
addition operator in distortion space is in fact non-linear
\citep{BJ02}, a detail that will be important, but manageable if we
avoid the $\approx 1\%$ of the sky where the lensing distortion is
strong.

Any weak lensing measurement maximizes the likelihood
of the measured $\bfe^{\rm obs}_k$ under some model for the shear planes
$\bfd_\ell$ (and perhaps the systematic error contribution simultaneously).  In
power-spectrum tomography, one varies the assumed power spectra of the
$\bfd_\ell$ to best match the observed covariances of the
$\bfe^{\rm obs}_k$.  In our case, we will assume that is possible to produce
some estimate $\hbfd_\ell$ of the distortion pattern without reference to
the background galaxy-shape information.  In practice this template distortion
would be developed from the galaxies identified near $z_\ell$ from the
photometric redshift data.  This could be as simple as convolving the
galaxy distribution with an isothermal halo model, or perhaps involve
a more sophisticated identification of groups and clusters. 
We allow that the template distortion is inaccurate to
some level, due to galaxy bias or other errors:
\begin{equation}
\langle \bfd_\ell \cdot \hbfd_\ell \rangle  =  \beta_\ell 
\langle d^2_\ell \rangle
\end{equation}
$\beta_\ell$ is a measure of the fidelity of our template, and is
related to the bias $b$ and correlation coefficient $r$ between
galaxies and mass, \eg\ as measured by \citet{Hoek02b}. 

To isolate the geometric term, we correlate the template distortion with
the measured distortion.  We wish to sum the measurements from all source
galaxies in some bin $s$ of redshift centered on $z_s$.  When the
shape noise is dominant, the sum which
is optimal in the sense of best $S/N$ on $g_{\ell s}$ is 
\begin{equation}
\label{xls}
X_{\ell s} \equiv {1 \over N_s} \sum_{k\in s} 
\bfe^{\rm obs}_k \cdot \hbfd_\ell(\bfx_k).
\end{equation}
Here $N_s$ is the number of source galaxies in bin $s$.
Under the model in \eqq{bfdk} the $X_{\ell s}$ will be
\begin{eqnarray}
\label{xlsmod}
X_{\ell s} & = & \sum_{\ell^\prime} R_{\ell\ell^\prime} g_{\ell^\prime s}
+ \langle \hbfd_\ell\cdot \bfd_{\rm sys}\rangle_s 
+ \langle \hbfd_\ell\cdot \bfe\rangle_s \\
\label{rll}
R_{\ell\ell^\prime} & \equiv & 
\langle
\hbfd_\ell \cdot \bfd_{\ell^\prime}
\rangle_s
\end{eqnarray}
Here the subscript $s$ on the average means that we are averaging over
the galaxies in source bin $s$, as in equation \ref{xls}, 
not over realizations of the distortion fields.  
For a given survey, the cosmology predicts the $g_{\ell s}$ values,
the templates are known, and the $R_{\ell\ell^\prime}$ are free
parameters.  The cosmological parameters $\{p_i\}$ are determined by
fitting the ${\bf X}=\{X_{\ell s}\}$ data vector to the model
(\ref{xlsmod}) with free parameters $\{p_i,R_{\ell\ell^\prime}\}$,
then marginalizing over the cross-correlations $\{R_{\ell\ell^\prime}\}$
to obtain confidence bounds for the $p_i$.
The intrinsic ellipticities and perhaps systematic errors act as
measurement noise on each $X_{\ell s}$.

We now assume that the intrinsic ellipticities $\bfe_k$ and the
systematic-error distortion $\bfd_{k, \rm sys}$ are independent of the
template.  We also assume that the distortions and templates are
uncorrelated between lens shells, so that $\langle
R_{\ell\ell^\prime}\rangle=0$ for $\ell\ne\ell^\prime$.  This last
assumption is true if the shells are much thicker than the correlation
length of the mass distribution; any correlations could be accommodated
in a more sophisticated analysis, or eliminated by applying high-pass
filters to the templates.
With these assumptions, the expectation value for the $X_{\ell s}$
when we average over realizations of the distortion fields is
\begin{equation}
\label{xlsmean}
\langle X_{\ell s} \rangle = \langle R_{\ell\ell} \rangle g_{\ell s}
= \beta_\ell \langle d^2_\ell \rangle g_{\ell s}.
\end{equation}

\subsection{Covariance Matrix and Likelihood}
In order to calculate probabilities for fits to the ${\bf X}$ vector
we need its covariance matrix ${\bf C}_x$.  From the definition (\ref{xls})
and the model (\ref{bfdk}) we can obtain the covariance in a
straightforward manner.  Note that when fitting to the model
(\ref{bfdk}) we will be considering only the shape noise and
systematic errors to be random variables.  The distortion planes, or
more precisely the cross-correlations $R_{\ell\ell^\prime}$, are
considered as free parameters in the fit, not random variables.
We will henceforth ignore the systematic distortion $\bfd_{\rm sys}$,
assuming that it is a minor contributor to the noise compared to the
shape noise $\bfe$.  We have already seen that the systematic errors
do not bias the observables if they are uncorrelated with the
templates.

We assume the $\bfe_k$ to be independent, with 
${\rm Var}(e_+)={\rm Var}(e_\times)\equiv\sigma_e^2$ the
variance in each component.  Intrinsic correlations of galaxy shapes
may slightly inflate our uncertainties, but should again be
uncorrelated with the $\hbfd_\ell$ and hence will not bias the fit.
In this case we have
\begin{eqnarray}
({\bf C}_x)_{\ell s \ell^\prime s^\prime}
 & = & {\rm Cov}(X_{\ell s} X_{\ell^\prime s^\prime})
 = \delta_{s s^\prime} {\sigma_e^2 \over N_s}
	\langle \hbfd_\ell \cdot \hbfd_{\ell^\prime} \rangle_s 
\nonumber \\*
& &  \left\{ \begin{array}{ll}
	= {\sigma_e^2 \langle \hat d^2_\ell \rangle \over N_s}
	& \ell=\ell^\prime \\
	\approx 0 & \ell\ne\ell^\prime
		       \end{array}
	\right.
\label{cx}
\end{eqnarray}
The $\ell\ne\ell^\prime$ cross-correlation is negligibly small if the
survey is much larger than the correlation length of the distortion
templates.

The distributions of the $\bfd_\ell$ and $\bfe_k$ are very
non-Gaussian.  For a survey with many source galaxies and an area much
larger than the distortion correlation length, however, the
distribution for $X_{\ell s}$ should approach Gaussian by the central
limit theorem.  The likelihood of an observed ${\bf X}$ vector
given a choice of model parameters is then given by the usual Gaussian
formula with covariance matrix ${\bf C}_x$.
With the likelihood, we may produce confidence
bounds on the parameters of the underlying geometry, once we
marginalize over the unknown $R_{\ell\ell^\prime}$.

\subsection{Fisher Uncertainties}
Given the likelihood function, the minimal uncertainties on the model
parameters may be derived from the Fisher matrix:
\begin{equation}
\label{fisher1}
F_{ij} = - \left\langle
{ \partial^2 {\cal L} \over \partial p_i \partial p_j }
\right\rangle
\end{equation}
where ${\cal L}\equiv \log L$.  The Fisher matrix for
the Gaussian case is well known, \eg\ \citet{TTH97}:
\begin{equation}
\label{tth}
{\bf F}_{ij} = {1 \over 2} {\rm
Tr}({\bf C}^{-1}{\bf C}_{,i}{\bf C}^{-1}{\bf C}_{,j})
+ \langle {\bf X}^T_{,i} \rangle {\bf C}^{-1} \langle {\bf X}_{,j}
\rangle.
\end{equation}
As usual the subscripts with commas denote derivatives.  This
simplifies considerably in our case, as ${\bf C}_x$ is diagonal and
independent of the parameters.

The model parameters are of two kinds:  the nuisance parameters
$\{R_{\ell\ell^\prime}\}$ and the cosmological parameters $\{p_i\}$.
We wish to split the Fisher matrix into submatrices ${\bf F}_{RR}$,
${\bf F}_{pR}$, and ${\bf F}_{pp}$.  Henceforth we will adopt the
convention that Greek indices (or $\ell$) range over the lens planes,
the $s$ index runs over source planes, and other  Latin
indices (except $\ell$) range over the cosmological parameters.  
Keeping in mind that it takes two Greek indices to specify one $R$
component, the three submatrices ${\bf F}_{RR}$,
${\bf F}_{pR}$, and ${\bf F}_{pp}$ are respectively
\begin{eqnarray}
\label{fishrr}
{\bf F}_{\alpha\beta\mu\gamma} & = & 
	\delta_{\alpha\mu} \sum_s {N_s \over \sigma^2_e 
	\langle \hat d^2_\alpha \rangle}
	g_{\beta s} g_{\gamma s} \\
 & = & \delta_{\alpha\mu} {N \over \sigma^2_e 
	\langle \hat d^2_\alpha \rangle}
	{\bf G}_{\beta\gamma} \\
{\bf F}_{i\alpha\beta} & = & 
	{N \over \sigma^2_e 
	\langle \hat d^2_\alpha \rangle}
	\sum_\gamma R_{\alpha\gamma}({\bf G}_i)_{\beta\gamma} \\
{\bf F}_{ij} & = & \sum_\alpha
	{N \over \sigma^2_e 
	\langle \hat d^2_\alpha \rangle}
	\sum_{\beta\gamma} R_{\alpha\beta}R_{\alpha\gamma}
	({\bf G}_{ij})_{\beta\gamma}.
\end{eqnarray}
The geometric factors are encoded in the matrices
\begin{eqnarray}
({\bf G})_{\alpha\beta} & \equiv & 
	\int d\!z_s {dn \over dz_s} g_{\alpha s}g_{\beta s} \\
({\bf G}_i)_{\alpha\beta} & \equiv & 
	\int d\!z_s {dn \over dz_s} g_{\alpha s}
	{dg_{\beta s}\over dp_i} \\
({\bf G}_{ij})_{\alpha\beta} & \equiv & 
	\int d\!z_s {dn \over dz_s}
	{dg_{\alpha s}\over dp_i}
	{dg_{\beta s}\over dp_j} 
\end{eqnarray}
The redshift distribution of sources $dn/dz$ is normalized to unit
integral, and $N$ is the total number of sources in the survey.  There
is no loss of information in moving to infinitesimal source-redshift
bins. 

\subsection{Prior Information}
We are interested in the covariance matrix ${\bf C}_p$ of the
geometric parameters after marginalization over the $R$ parameters.
If $L$ is the number of lens planes in the model, there are $L^2$ free
$R_{\ell\ell^\prime}$ being marginalized, and after this
marginalization the constraints on the cosmological parameters are
relatively weak for envisioned surveys.

There is, however, additional information about the $R$ values that
has not been incorporated into the model of \eqq{xlsmod}, and hence
not in the Fisher matrix of \eqq{fishrr}.  We expect that the
$R_{\ell\ell^\prime}$ are near zero for $\ell\ne\ell^\prime$ because
the lens planes are uncorrelated.  We incorporate this knowledge with
a prior probability on the $R$ values.  If the prior is Gaussian, then
we may add the inverse of the $R$ covariance matrix to the ${\bf
  F}_{RR}$ component of the Fisher matrix.

Note that until this point we have had no use for an {\em ensemble} of
mass distributions; all of the quantities in the solution and Fisher
matrix for the cosmological parameters make use of just the
$R_{\ell\ell^\prime}$ for the single
realization of the mass distribution that exists in our survey field.
It is only in assembling a prior on these values (and in estimating
typical values in \S\ref{marginalize}) that we make use of the
ensemble properties of the mass distribution.  This is why the
sample variance contribution is unimportant in the uncertainties of the
cross-correlation method.

In the Appendix we show that the covariance matrix for
$R_{\ell\ell^\prime}$ is nearly diagonal.  The diagonal terms are
\begin{equation}
\label{varrr}
{\rm Cov}(R_{\alpha\beta}R_{\alpha\beta}) =
\langle \hat d^2_\alpha \rangle
\langle d^2_\beta \rangle
{\Omega_{\alpha\beta} \over \Omega}.
\end{equation}
Here $\Omega$ is the total solid area of the survey and
$\Omega_{\alpha\beta}$ is a measure of the area of coherence of the
two distortion fields $\bfd_\alpha$ and $\bfd_\beta$.  For redshift
ranges of interest, $\Omega_{\alpha\beta}\lesssim1\,{\rm arcmin}^2$.
The only off-diagonal terms in the covariance matrix for the $R$ values 
are ${\rm Cov}(R_{\alpha\beta}R_{\beta\alpha}).$
To simplify the following algebra,
we take the covariance matrix for our prior on $R$ to have diagonal
elements that are twice as large as \eqq{varrr} and drop the
covariance between $R_{\alpha\beta}$ and $R_{\beta\alpha}$.
This prior distribution is less restrictive than is the full
covariance matrix, so we are at liberty to make this choice.

As a cautionary step we do not make use of the prior information on
$R_{\alpha\alpha}$.  This is because $R_{\alpha\alpha}$ appears in the
expectation value $\langle X_{\alpha s}\rangle$ in \eqq{xlsmean}, and
hence prior assumptions may bias the fit for the cosmological
parameters that drive the $g_{\alpha s}$ in this fit.  

With this simplified, diagonal prior for the $R$ values,
the Fisher matrix for the system is altered from \eqq{fishrr} as
\begin{eqnarray}
\label{newfishrr}
{\bf F}_{\alpha\beta\mu\gamma}
& \rightarrow &
\delta_{\alpha\mu} {N \over \sigma^2_e 
	\langle \hat d^2_\alpha \rangle}
\left[
  {\bf G}_{\beta\gamma} 
 + \delta_{\beta\gamma} (1-\delta_{\alpha\beta})
 { \sigma^2_e \over 2n \Omega_{\alpha\beta} \langle d^2_\beta \rangle
}
\right] \nonumber \\
 & \equiv & 
\delta_{\alpha\mu} {N \over \sigma^2_e 
	\langle \hat d^2_\alpha \rangle}
(  {\bf G} + {\bf P}_\alpha )_{\beta\gamma}
\end{eqnarray}
${\bf P}_\alpha$ is simple, having only diagonal elements and a zero
at the $\alpha$ element of the diagonal.  Here $n=N/\Omega$ is the
sky density of source galaxies.

\begin{widetext}
\subsection{Marginalization}
\label{marginalize}
We now return to the marginalization over the $R_{\alpha\beta}$ to
obtain a covariance matrix $C_p$ for the cosmological parameters.
Using a common matrix identity, we obtain
\begin{equation}
({\bf C}_p^{-1})_{ij} =
\left({\bf F}_{pp} 
- {\bf F}_{pR} {\bf F}_{RR}^{-1} {\bf F}_{pR}^T \right)_{ij}
  =  
	{N \over \sigma^2_e} \sum_\alpha
	{1 \over \langle \hat d^2_\alpha \rangle} 
 \sum_{\beta\gamma} R_{\alpha\beta}R_{\alpha\gamma}
	\left( {\bf G}_{ij} - {\bf G}_i^T ({\bf G}+{\bf
	  P}_\alpha)^{-1} {\bf G}_j 
	\right)_{\beta\gamma}.
\label{cp1}
\end{equation}
This gives the Fisher uncertainties on cosmological parameters for a
survey over a sky with given distortion field and hence given $R$
values.  We next average over realizations of the mass distribution in
the Universe, which requires
that we calculate
\begin{equation}
\label{rrexpect}
\langle R_{\alpha\beta}R_{\alpha\gamma} \rangle  =  
{\rm Cov}(R_{\alpha\beta}R_{\alpha\gamma})
 + \langle R_{\alpha\beta}\rangle\langle R_{\alpha\gamma}\rangle 
  \equiv  \delta_{\beta\gamma} \left[
  \langle\hat d^2_\alpha\rangle
  \langle d^2_\beta\rangle {\Omega_{\alpha\beta} \over \Omega}
 + \delta_{\alpha\beta} 
   \langle R_{\alpha\alpha}\rangle^2 \right].
\end{equation}

Combining Equations~(\ref{xlsmean}), (\ref{cp1}), and
(\ref{rrexpect}), we obtain the expected (inverse) covariance matrix
for the cosmological parameters:
\begin{equation}
\label{cp2}
({\bf C}_p^{-1})_{ij}  = 
	{N \over \sigma^2_e} \sum_\alpha \left[
	\beta_\alpha^2 \langle \hat d^2_\alpha \rangle 
	\left( {\bf G}_{ij} 
	- {\bf G}_i^T ({\bf G}+{\bf P}_\alpha)^{-1} {\bf G}_j
	\right)_{\alpha\alpha} 
 +  \sum_\beta
	\langle d^2_\beta \rangle 
	{\Omega_{\alpha\beta} \over \Omega}
	\left( {\bf G}_{ij} 
	- {\bf G}_i^T  ({\bf G}+{\bf P}_\alpha)^{-1} {\bf G}_j
	\right)_{\beta\beta}
\right]
\end{equation}
\end{widetext}
We note first that there is no ``sample variance limit'' to the
cross-correlation cosmography measure of the cosmological
parameters---uncertainties scale as $1/\sqrt N$ as the
density of sources becomes large.  Sample variance typically arises
when one does 
not measure enough independent patches of the density field to
adequately constrain its power spectrum.  In our application, however,
we are not using the power spectrum to measure cosmology, we instead
are just cross-correlating the observations with whatever mass
distribution the Universe gives us.  Hence there is no sample variance
contribution. 

The left-hand term in \eqq{cp2} arises purely from the shape noise in
the measured field that is being cross-correlated with the templates,
while the right-hand term arises because the distortion from one lens
slice acts as a correlated noise source for all the $X_{\ell s}$.
The right-hand term has an amplitude $\approx L/N_c$
relative to the left-hand term, where $L$ is the number of lens planes
and $N_c=\Omega/\Omega_{\alpha\beta}$ is the number of independent
cells or patches of the distortion fields.  In almost any large survey
we should be able to choose the lens-plane width $\Delta z_\ell$ to be
sufficiently narrow that the $g_{\ell s}$ do not vary significantly
within the bin, and still have $L/N_c\ll 1$.  In other words we will
usually have more than enough independent patches to decorrelate the
different lens planes.

In the regime where the left-hand term dominates, the parameter
accuracy is essentially independent of the foreground binning, as long
as the fidelity $\beta_\ell$ remains high and the shells remain
uncorrelated.

\section{Application to Candidate Surveys}
\subsection{Cosmological Parameters}
Any parameter that influences
 growth factor $a(t)$ or curvature of the Universe will have some
 effect on the ${\bf X}$ data via $g_{\ell s}$; in particular we will
 be interested in the present-day matter and energy densities
 $\Omega_m$ and $\Omega_X$, and the parameters $w_0$ and $w_a$ that
 approximate the equation of state of the dark energy via
 \citep{Linder03} 
\begin{equation}
\label{wadef}
P_X = [w_0 + w_a(1-a)]\rho_X.
\end{equation}
We will henceforth assume a flat Universe $\Omega_X = 1-\Omega_m$, 
in which case we have \citep{Linder03}
\begin{eqnarray*}
r[\chi(z)]  & = &  \chi(z) = {c \over H_0}\int_0^z d\!z^\prime
\left[ \Omega_m(1+z^\prime)^3 + \right. \\
& & \left.(1-\Omega_m) (1+z^\prime)^{3(1+w_0+w_a)}
	e^{-3w_az^\prime/(1+z^\prime)} \right]^{-1/2}
\end{eqnarray*}
Unless otherwise noted, we will calculate the sensitivity to
departures from the canonical $\Lambda$CDM model ($\Omega_m=0.3,
w_0=-1, w_a=0$).

\subsection{Mass Variance}
The distortion power generated by a mass slice at $z_l$ and the
coherence areas $\Omega_{\alpha\beta}$ can be
estimated by applying the formulae in the Appendix to a model
for the non-linear evolution of the power spectrum.
We reiterate that errors in this model
will affect only our forecasted uncertainty, not the derived
cosmological parameters.  

We use the nonlinear mass power spectrum given by the fitting 
formula of \citet{Pea96}, with implementation for weak lensing as
described in \citet{JS97}. We need this only for 
the fiducial $\Lambda$CDM model, for which the differences between
different fits to N-body simulations are small \citep{Sm02}. 
The integrals over the power spectrum
are cut off for $k>2\pi/(50\,{\rm kpc})$; this is a crude way of
accounting for the fact that we will have to exclude the central
regions of galaxy clusters, where the lensing is strong and where the
light of the foreground galaxies may preclude observation of the
background galaxies.  The results are not sensitive to the choice
of the high-$k$ cutoff. We will assume that the estimated templates
$\hbfd_\ell$ have the same variance as the true distortion fields
$\bfd$, but with fidelity $\beta_\ell=0.8$.  The uncertainties on
cosmological parameters will scale as $\beta_\ell^{-1}$.

\begin{figure}[t]
\plotone{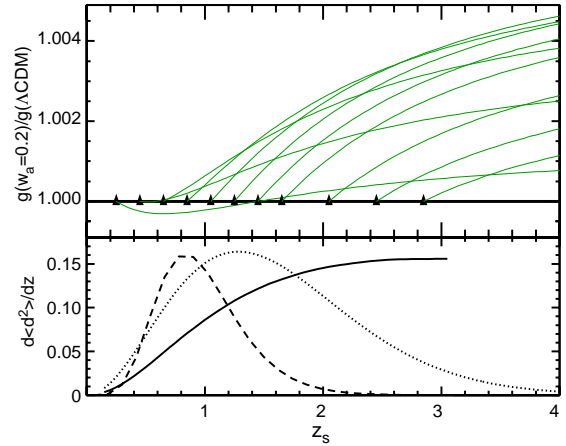}
\caption[]{\small
The top panel shows the fractional change in the geometric factor
$g_{\ell s}=(\chi_s-\chi_\ell)/\chi_s$ when we shift from a pure
$\Lambda$CDM 
Universe to one with $w_a=0.2$; this should be discernible at
$1\sigma$ significance in the SNAP survey.  The horizontal axis is
$z_s$ and each 
line corresponds to a different $z_\ell$.  The triangle at the end of
each line marks $z_\ell$.  Because the mass normalization in each lens
slice is free to vary, vertical shifts of each line carry no
cosmological information, so we align them all to be unity at
$z_\ell$.  The cosmological information is carried in the departures
of each line from horizontal; these departures are small, amounting to
only a few parts in $10^{-3}$ change in the induced background
distortion.  The smallness of this signal implies that the calibration
of the distortion measurement and the source redshifts must be
accurate to roughly a part in $10^3$.
The lower panel plots the assumed source redshift distribution (dotted
line) and the expected distortion variance per unit redshift (solid
line) using estimated non-linear power spectra.  The dashed line shows
the relative contribution of different lens planes to the constraint
on $w_a$. 
}
\label{dgdwa}
\end{figure}

The bottom panel of Figure~\ref{dgdwa} plots the strength of lensing
distortion vs redshift, $d\langle d^2\rangle/dz$ that arises from
this model.  For the coherence angle we use
\begin{equation}
\label{oab}
\Omega_{\alpha\beta} = (0.24\,{\rm arcmin})^2 
	(\chi_\alpha \chi_\beta)^{-0.8},
\end{equation}
which is within $\sim25\%$ of the more carefully calculated values.
The coherence area has only a minor impact on the cosmological
uncertainties. 

\subsection{Redshift Distribution}
We adopt the common guess for the redshift distribution of faint
galaxies,
\begin{equation}
{dn \over dz} \propto z^2 \exp\left[-(z/z_0)^{1.5}\right],
\end{equation}
and will select $z_0$ and the overall density $n$ to crudely mimic the
expectations of several future surveys.

\subsection{Parameter Forecasts}
Table~\ref{surveys} lists crudely approximated parameters for three
possible weak-lensing surveys:  the CFHT Legacy Survey, just beginning 
and expected to take 5 years to cover
$\approx200\,{\rm deg}^2$ of sky; a possible deep multicolor lensing
survey covering $\sim10\%$ of the sky with the proposed Large Synoptic
Survey Telescope (LSST)\footnote{\tt www.dmtelescope.org/dark\_home.html},
and a 400~deg$^2$ weak lensing survey with the proposed
Supernova Acceleration Probe (SNAP)\footnote{\tt snap.lbl.gov}. The
orbiting SNAP focal plane should obtain significantly higher source
galaxy densities than the ground-based observatories, with
consequently deeper median redshift.

In evaluating the Fisher matrix, we assume that $\sigma_e=0.3$ and
that $\beta_\ell=0.8$.  Cosmological parameter uncertainties will
scales as $\sigma_e/\beta$.  We use $\Delta z_\ell=0.2$
for the thickness of the lens slices.  This choice has very little
impact upon the parameter constraints because the first term dominates
\eqq{cp2}, \eg\ changing to $\Delta z_\ell=0.1$ for the SNAP case
alters the parameter constraints by $\lesssim 1\%$.

\begin{figure}
\plotone{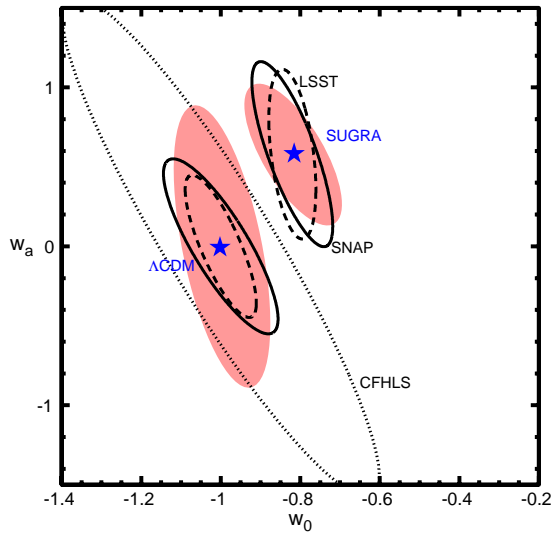}
\caption[]{\small
Fisher uncertainty ellipses for dark energy parameters derived from
three candidate weak lensing surveys are plotted.  All ellipses are
68\% confidence two-dimensional regions ($\Delta\chi^2=2.3$) after
application of a Gaussian prior on $\Omega_m$ with $\sigma=0.03$ and
marginalization over $\Omega_m$.  The solid, dashed, and dotted
contours are for SNAP, LSST, and CFHLS surveys, respectively.
Survey parameters are listed in
Table~\ref{surveys} and a flat Universe is assumed.  Fiducial models
for both $\Lambda$CDM and a supergravity-inspired model are plotted
as stars, and the shaded regions are the expected
constraints from the SNAP Type Ia supernova measurement plus
$\Omega_m$ prior (E. Linder, private communication).  Unlike the weak
lensing contours, the SN contours include the estimated effects of the
dominant systematic errors.}
\label{ellipses}
\end{figure}

For each candidate survey we evaluate the cosmological-parameter
covariance matrix ${\bf C}_p$ using \eqq{cp2}.  The resultant matrix
is highly degenerate in one direction, so we adopt a Gaussian prior
distribution on $\Omega_m$ with $\sigma=0.03$.  The constraints in the
$w_0-w_a$ plane after applying this prior and marginalizing over
$\Omega_m$ are plotted in Figure~\ref{ellipses} for the three
surveys.  Table~\ref{surveys} also lists the 1-$\sigma$ uncertainties
in $w_0$ and $w_a$ (assuming in each case marginalization over all
other parameters) and the correlation coefficient between these
two parameters.

For both the SNAP and LSST surveys,
the constraints on the dark energy equation of state and its evolution
are quite strong, comparable to or tighter than those of any proposed
experiment of which we are aware.   The 1-$\sigma$ uncertainties on
$w_a$, even after marginalizing over all other parameters, are
$\sigma_{w_a}\approx0.3$ in both cases.  
In terms of the commonly used parameter 
$w^\prime\equiv [dw/d\ln(1+z)]_{z=1}=w_a/2$
we have $\sigma_{w^\prime}\approx0.15$.

\subsection{Dependence of Constraints on Survey Parameters}
The Fisher uncertainties from the weak lensing survey will scale as
$f_{\rm sky}^{-1/2}$.  Because the prior on
$\Omega_m$ is independent of $f_{sky}$, the marginalized error ellipses do
not quite scale as $f_{\rm sky}^{-1/2}$, but this is a valid
approximation in the neighborhood of the canonical survey parameters.

Increased survey depth leads to both greater source density $n$ and
greater median redshift $z_{\rm med}$.  Parameter uncertainties scale
as $n^{-1/2}$ (again not quite because of the prior), and hence unlike
power-spectrum tomography, there is no breakpoint at which cosmic
variance limits a survey of a given size.  The additional depth
is also a benefit:  changing $z_{\rm med}$ from 1.0 to 1.5 decreases the
error on $w_a$ by 23\%, if the galaxy density and survey area are held
fixed.   In other words, the increased depth makes each galaxy about
1.5 times more valuable.

The lower panel of Figure~\ref{dgdwa} illustrates the contribution of
different lens planes to the constraint on $w_a$ for the SNAP survey.
Most of the information comes from lens planes near $z=1$,
but if a deeper source distribution were available, lens planes at
higher $z$ would continue to add significant information.  It would be
particularly interesting to see how this behavior is altered with
inclusion of the cosmic microwave background as a source screen at
$z\simeq 1000$. 

Combining these various scalings, the overall constraint on evolution
of the dark energy equation of state will scale roughly as
\begin{equation}
\sigma_{w_a} \propto \left( { \sigma_e^2 \over n f_{\rm sky}\beta^2 }
\right)^{0.5} z_{\rm med}^{-0.6}
\end{equation}

\subsection{Comparison with Jain \& Taylor}
Our parameter constraints can be compared with those obtained by
\citet{JT} who used a simplified implementation of 
the cross-correlation approach. The measurement suggested by them was the
tangential shear around foreground halos, identified using galaxy
groups and clusters. They assumed that essentially all halos out
to $z=1$ with mass $> 4\times10^{13} M_\odot$ could be identified
this way. For a given lens slice they used source galaxy shapes on 
only $\simeq 10\%$ of the sky.  This approach corresponds to 
taking a particularly simple construction of the template shear map. 
Other simplifications made by Jain \& Taylor were that 
intrinsic ellipticity shot noise was taken to be the only source of
error and only two bins in source redshift were used for each lens slice. 
We find that the method presented here can improve parameter 
constraints, primarily due to the use of more than two redshift slices. 
However our results for $\sigma_w$ and $\sigma_{w_a}$ are close to 
those of Jain \& Taylor because the size of the single-parameter
marginalized errors is, for the large surveys
considered, controlled by the prior for $\Omega_m$.

\begin{deluxetable*}{lcccccc}
\tablewidth{0pt}
\tablecaption{Future Weak Lensing Surveys and Dark Energy Constraints}
\tablehead{
\colhead{Survey} &
\colhead{Median $z$} &
\colhead{Galaxy Density} &
\colhead{$f_{\rm sky}$} &
\colhead{$w_0$ Error} &
\colhead{$w_a$ Error} &
\colhead{$w_0-w_a$ Correlation} \\
 & & \colhead{(arcmin$^{-2}$)} & & & &
}
\startdata
CFHLS & 1.0 & 30 & 0.005 & 0.26 & 1.04 & -0.92\\
LSST & 1.0 & 30 & 0.1 & 0.06 & 0.30 &  -0.80\\
SNAP & 1.5 & 100 & 0.01 & 0.10 & 0.36 & -0.83
\enddata
\label{surveys}
\end{deluxetable*}

\section{Systematic Errors}
The cross-correlation cosmography technique is essentially insensitive
to several of the largest systematic error sources in power-spectrum
tomography, namely residual PSF distortion and errors in calculation
of the non-linear power spectrum.  The technique does, however, place
stringent demands on the distortion calibration and photometric
redshift accuracy of the weak lensing survey.  This is illustrated in
Figure~\ref{dgdwa}.  The top panel plots the fractional change in
$g_{\ell s}$ when we move from a $\Lambda$CDM universe to one with
$w_a=0.2$.  According to the Fisher calculation, this should be
detectable at 1$\sigma$ significance in the SNAP survey.

\subsection{Demands on Distortion Calibration}

Each line in the plot shows $g_{\ell s}$ vs $z_s$ for one choice of
$z_l$.  The ``signal'' in the cross-correlation cosmography method is
the departure of these curves from horizontal lines.  A pure vertical
shift of any line will be degenerate with a change in the fidelity
$\beta_\ell$ of the template in lens shell $\ell$, hence the
cosmological information is in the slope/curvature of these lines.  We
see that the minimum detectable cosmological signature corresponds to
a change of $\approx2$ parts in $10^3$ of $g$, and hence an equivalent
change in the measured distortion vs $z_s$.  We may immediately
conclude that {\em exploitation of this technique to constrain $w_a$
requires that the distortion calibration be constant to 1 part in
$10^3$ over all measured redshifts.}  This will be a significant
technical challenge.  \citet{Hirata} demonstrate that present
algorithms have calibration errors of 1--10\%, so at least an
order-of-magnitude improvement is required.  \citet{Hirata} and
\citet{BJ02} propose methods to improve calibration accuracy, but
these have not yet been demonstrated on real data.  Furthermore, most
methodologies assume that the distortion is weak, \ie\ they ignore
induced changes to galaxy shape that are of order $d^2$ or
higher.  The non-linear effects of even ``weak'' distortions will have
to be accounted for to second or even third order to reach the
calibration accuracy of $10^{-3}$.  Regions where the lensing is
strong will clearly have to be avoided.

To estimate the demands of this requirement on the data reduction
methodology, we note that the conversion of observed ellipticities to
pre-seeing elliptiticities typically involves the factor
$(1+r_\star^2/r^2)$, where $r$ is the angular radius of the
(pre-seeing) galaxy and $r_\star$ is the PSF radius.  A good weak
lensing methodology makes use of galaxies as small as $r\approx
r_\star$; in this case, both $r$ and $r_\star$ must be known to better
than 1 part in $10^3$ in order to obtain a distortion calibration
accurate to 1 part in $10^{-3}$.  If $r$ were independent of $z$, then
we would have redshift-independent calibration errors, which do not
affect the cross-correlation method.  But fainter, more distant
galaxies are typically smaller in angular diameter, so errors in
$r_\star$ will couple to the cosmography measurement.

\subsection{Demands on Redshift Determinations}
Our formalism can be adapted to deal with the {\em random errors}
in photometric redshift measurements, but it will still be crucial to
minimize {\em biases} in the photo-$z$ estimates.  The quantity
$g_{\ell s}$ scales slightly less than linearly with the redshifts
$z_s$ and $z_\ell$.  It is therefore clear that mis-estimates of the
mean photometric redshift of a few parts in $10^3$ would overwhelm the
signal of a $w_a=0.2$ cosmology.
Hence {\em photometric redshifts must be
accurate to $\approx10^{-3}$ in $\log(1+z)$.}  This again is at least
one order of magnitude beyond the present state of the art.

It is important to realize first that this is not the requirement on
the accuracy of {\em each} measured photometric redshift, but rather a
requirement on the {\em bias} of a collection of photometric redshift
estimates.   Second, we are not required to use all the galaxies in
the image.  One would likely choose to exclude from the analysis
galaxies whose colors make assignment of a photometric redshift
particularly uncertain or ambiguous.

\subsection{Practical Issues}
If the required accuracy of distortion and redshift calibration cannot
be achieved, one could introduce additional free parameters in the
model to represent calibration errors.  These terms would be purely
functions of $z_s$, and hence in theory distinguishable from the
cosmological signals, which couple $z_\ell$ and $z_s$ through $g_{\ell
s}$.  The constraints on dark energy would be degraded, to an extent
that is calculable with further Fisher analysis.

We note that if the random error in a typical photometric redshift is
$\sigma_z \approx 0.03(1+z)$, then it will take a spectroscopic survey
of $\approx10^3$ galaxies in order to check that the mean error (bias)
is below $10^{-3}(1+z)$.  Repeating this bias check in bins of
redshift from 0--3 would thus require a sample of $10^4$ or $10^5$
spectroscopic redshifts to compare with the photometric redshifts.
Redshift surveys of this size to limiting magnitudes of 24 or 25 will
be feasible:  indeed the DEEP2
survey\footnote{http://deep.berkeley.edu} is already well on its way
to its goal of 65,000 spectra to $I_{AB}<24.5$.

Given the extreme demands that cross-correlation cosmography will
place on the calibration of both the lensing distortion and photo-$z$
calibrations, there will be a significant advantage to space-based
observations.  Ground-based analyses must deal with a
constantly-varying PSF and atmospheric transmission function which
will make it more difficult to achieve these accurate calibrations.
A space-based platform will have the additional advantages of thermal
stability, a much sharper PSF, and the possibility of using near-IR
filter bands to improve the photo-$z$ accuracy.  

\section{Conclusion}
We have presented a formalism for implementing the idea of \citet{JT} to
constrain cosmology by tracing the dependence of induced distortion on
background redshift for a fixed foreground mass template.
Cross-correlating the background distortion with a series of
foreground mass templates has significant advantages over
power-spectrum tomography, particularly its immunity to spurious
distortion signals, and the ability to use non-linear lensing
structures without having to model accurately the non-linear evolution
of matter in the Universe.  We believe this formalism makes use of all
the information available in all lens-source pairs in a nearly optimal
fashion; we find that a survey with median redshift $\approx 1.5$ and
100~galaxies per square arcminute can constrain the dark energy
parameters to $\sigma_{w_0}\approx 0.01 f_{\rm sky}^{-1/2}$ and
$\sigma_{w_a}\approx 0.035 f_{\rm sky}^{-1/2}$ after application of a
practical prior on $\Omega_m$.

Realization of the full potential of the cross-correlation cosmography
method will require that techniques for the calibration of lensing
distortion and photometric redshift be improved by at least an order
of magnitude from present state of the art.  There are no known
fundamental barriers to this, but it will not be easy.  
We discuss ways of fitting for parameters in the calibration from the 
data. To reduce possible biases in the photometric redshifts to an 
acceptable level would require spectroscopic redshifts for 
$10^4-10^5$ galaxies over the redshift range used in the analysis. 
An orbiting
observatory may be preferred for obtaining the precision required for
our method due to its
greater photometric and optical stability, and access to the near
infrared.

The potential accuracy of cross-correlation cosmography on the
equation-of-state time variation compares well to the expected
precision of power-spectrum tomography.  A formalism for full
utilisation of the power-spectrum tomography information has not yet
been published, and none of the few published investigations of the
effects of time-dependent dark energy use \eqq{wadef} for the equation
of state.  The assumed priors and confidence levels of published
estimates also vary greatly, so only crude comparisons can be made.
\citet{Hu02b} estimates an error of $\approx 0.05 f_{\rm sky}^{-1/2}$
on $w^\prime$ from tomography confined to the linear regime. The plots
of \citet{BvW03} suggest $\approx 0.03 f_{\rm sky}^{-1/2}$ from a
non-tomographic analysis deep into the non-linear regime.  It should
ultimately be possible to use all the information encoded in the weak
lensing---cosmography, growth function, and cross-correlation with
foreground structures---to provide constraints stronger than those we
forecast here.  By cancelling the growth information, however, we
eliminate the systematic errors that might arise from mis-calculation
of the theoretical non-linear spectrum, and from {\em additive}
contamination of the distortion field by systematic effects.  The
penalty for cancelling the growth factor is that the signal is
partially cancelled as well, leaving the cross-correlation method more
susceptible to {\em multiplicative, redshift-dependent} errors in the
distortion field that might arise from PSF effects.

Further improvements to the cross-correlation cosmography methodology
are worth investigation.  The cosmic microwave background can serve as
an additional source plane at $z_s=1000$, and its shear pattern can be
cross-correlated with all the lens planes to provide a $S/N$
improvement and greater redshift leverage.  One could also make use of
magnification information as well as shear in the background galaxy
samples, which would improve the $S/N$ and serve as a useful cross-check
(\eg\ \citet{Jain02}). Finally, in this study we have not developed a detailed
method to obtain template distortion maps from the foreground galaxy
distribution. A detailed study that includes the redshift dependence of the 
fidelity of the template would be useful. 

\acknowledgements
GMB is supported in this work by grant AST-9624592 from the National Science
Foundation. 
BJ acknowledges the Aspen Center for Physics where part of this
work was done and support from NASA grant NAG5-10923.
We thank Wayne Hu, Andy Taylor, Jun Zhang, Masahiro Takada and Michael
Jarvis for helpful conversations.

\newpage
\appendix
\section{Covariance Matrix of $R_{\alpha\beta}$}
\label{pspec}
At several points in the analysis we require the covariances of the
slice-to-slice distortion correlations 
$R_{\alpha\beta}\equiv \langle \hbfd_\alpha \cdot \bfd_\beta \rangle$.  Under
the assumption that $\bfd_\beta$ is completely uncorrelated with both
$\bfd_\alpha$ and $\hbfd_\alpha$, it is clear that the only
non-vanishing elements of the covariance matrix are of the form
${\rm Var}(R_{\alpha\alpha}),$
${\rm Var}(R_{\alpha\beta}),$
and ${\rm Cov}(R_{\alpha\beta} R_{\beta\alpha}).$
We first calculate ${\rm Var}(R_{\alpha\beta})$ for $\alpha\ne\beta$:
\begin{eqnarray}
\label{rrexpect2}
\langle R_{\alpha\beta}R_{\alpha\beta} \rangle & = & 
	{1 \over \Omega^2} \int\int d^2\!r \,d^2\!r^\prime \,
	\left\langle [\hbfd_\alpha({\bf r})\cdot\bfd_\beta({\bf r})]
         [\hbfd_\alpha({\bf r^\prime})\cdot\bfd_\beta({\bf
	r}^\prime)]
	\right\rangle \\
 & = &
{2 \over \Omega} \int d^2r \left[
	\xi_{\alpha+}(r) \xi_{\beta+}(r) +
	\xi_{\alpha-}(r) \xi_{\beta-}(r)
  \right] \\
 & = & {2 \over \Omega} \int {d^2\ell \over (2\pi)^2}
	P_\alpha(\ell) P_\beta(\ell).
\end{eqnarray}
$\Omega$ is the solid angle of the survey, and $P_\alpha(\ell)$ and
$P_\beta(\ell)$ are the power spectra of the lensing convergence
corresponding to the distortion fields $\hbfd_\alpha$ and
$\bfd_\beta$, respectively, at spherical harmonic $\ell$.  We have
made use of the shear correlation
functions $\xi_\pm$ as defined for example in \citet{Sch02}, who also
give these correlation functions in terms of the convergence power
spectrum.  We may also put the distortion variance in terms of the
convergence power spectrum:
\begin{equation}
\langle d^2_\beta \rangle =
{4 \over (2\pi)^2} \int d^2\ell P_\beta(\ell).
\end{equation}
We can now express the covariance straightforwardly as
\begin{eqnarray}
\label{omegaab}
\langle R_{\alpha\beta}R_{\alpha\beta} \rangle & = & 
\langle \hat d^2_\alpha \rangle
\langle d^2_{\beta} \rangle
{ \Omega_{\alpha\beta}  \over \Omega} \\
 \Omega_{\alpha\beta} & = &
{ \pi^2 \over 2} { \int d^2\ell\, P_\alpha(\ell)P_\beta(\ell) \over
\int d^2\ell\, P_\alpha(\ell) \int d^2\ell\, P_\beta(\ell) }.
\end{eqnarray}
The quantity $\Omega_{\alpha\beta}$ gives the solid angle over which
the two distortion fields maintain mutual coherence.  We will
make the approximation that the template field $\hbfd_\alpha$ has a
power spectrum with the same shape as the actual field $\bfd_\alpha$,
so that $\Omega_{\alpha\beta}$ has the same value regardless of
whether we are correlating the templates or the real distortion, \eg
$\Omega_{\beta\alpha}=\Omega_{\alpha\beta}.$ 

If the lens shell thickness $\Delta\chi$ satisfies
$\Delta\chi\ll\chi_\ell$, then we have a simple relation between 
the convergence power spectrum and the spectrum $P_{3d}$ of mass
fluctuations in three dimensions:
\begin{equation}
P_\beta(\ell) = { 9 H_0^2 \Omega_m^2 \over 4 a^2(\chi_\beta) c^4}
P_{3d}\left({\ell \over \chi_\beta}, \chi_\beta\right) \Delta\chi
\end{equation}

${\rm Cov}(R_{\alpha\beta} R_{\beta\alpha})$ is also non-vanishing; it
differs from ${\rm Var}(R_{\alpha\beta})$ in that cross-correlations
between $\hbfd_\alpha$ and $\bfd_\alpha$ are required.  If the
template is well correlated with the actual mass then we expect
\eqq{omegaab} to describe the covariance up to a factor near unity,
\ie\ $R_{\alpha\beta}$ will be highly correlated with
$R_{\beta\alpha}$.  We do not need this expression in detail.

The final non-vanishing term is (again ignoring the distinction
between template and actual distortion)
\begin{equation}
{\rm Var}(R_{\alpha\alpha}) =
{1 \over \Omega} \int d^2r \left \langle 
\left[d_\alpha^2(0)-\langle d^2\rangle \right]
\left[d_\alpha^2({\bf r})-\langle d^2\rangle \right]
\right\rangle,
\end{equation}
which clearly involves the four-point correlation functions of the
shear field, and hence depends upon the degree of non-Gaussianity in
the mass distribution.  \eqq{omegaab} will be correct up to a factor
of order unity, which we could absorb into the definition of
$\Omega_{\alpha\alpha}$.  This term does not
significantly affect the Fisher matrix.

\end{document}